\documentclass{article}
\usepackage{amssymb}
\usepackage{amsmath}
\usepackage{amsfonts}
\usepackage{graphicx}

\begin{document}

\begin{center}
{\huge Chance in the Everett interpretation}

\bigskip

{\Large Simon Saunders\bigskip }
\end{center}

\noindent It is unanimously agreed that statistics depends somehow on
probability. But, as to what probability is and how it is connected with
statistics, there has seldom been such complete disagreement and breakdown
of communication since the Tower of Babel (Savage [1954 p.2]).

\begin{abstract} The notion of objective probability or chance, as a physical trait of the world, has proved elusive; the identification of chances with actual frequencies does not succeed. An adequate theory of chance should explain not only the connection of chance with statistics, but with degrees of belief, and more broadly the entire phenomenology of (seemingly) chance events and their measurement. Branching structure in the decoherence-based many worlds theory provides an account of what chance is that satisfies all these desiderata, including the requirement that chance involves uncertainty.\footnote{%
This is a lightly edited reprint (with added footnotes and a new appendix) of `Chance in the Everett interpretaiton', in \textit{Many Worlds? Everett, quantum theory and reality}, S. Saunders, J. Barrett, A. Kent, and D. Wallace (Oxford 2010).}  
\end {abstract}

\noindent For present purposes I take Everettian quantum \
mechanics (EQM) to be the unitary formalism of quantum mechanics divested of
any probability interpretation, but including decoherence theory and the
analysis of the universal state in terms of emergent, quasiclassical
histories, or \textit{branches}, along with their branch amplitudes, all
entering into a vast superposition. I shall not be concerned with arguments
over whether the universal state does have this structure; those arguments
are explored elsewhere (see e.g. Saunders et al [2010 Part 1,2]). (But I\ shall,
later on, consider the mathematical expression of this idea of branches in
EQM in more detail.)

My argument is that the branching structures in EQM, as quantified by branch
amplitudes (and specifically as ratios of squared moduli of branch
amplitudes) play the same roles that chances are supposed to play in
one-world theories of physical probability. That is, in familiar theories,
we know that

\begin{enumerate}
\item[(i)] Chance is measured by statistics, and perhaps, among observable
quantities, only statistics, but only with high chance.

\item[(ii)] Chance is quantitatively linked to subjective degrees of belief,
or credences: all else being equal, one who believes the chance of $E$ is $p$
will set his credence in $E$ equal to $p$ (the so-called `principal
principle').

\item[(iii)] Chance involves uncertainty; chance events, prior to their
occurrence, are uncertain.
\end{enumerate}

\noindent Those seem the most important of the chance-roles.

My claim is that exactly similar statements can be shown to be true of
branching in EQM. In the spirit of Saunders [2005] and Wallace [2006], we
should conclude these branching structures \textit{just are} chances, or
physical probabilities. This is the programme of `cautious functionalism',
to use Wallace's term.

The argument continues: this identification is an instance of a general
procedure in the physical sciences. Probabilities turn out to be functions
of branch amplitudes in much the same way that colours turn out to be
functions of electromagnetic frequencies and spectral reflectancies, and
heat and temperature turn out to be functions of particle motions in
classical statistical mechanics -- and in much the same way that sensible
perceptions (of the sort relevant to measurement and observation) turn out
to be functions of neurobiological processes.

Just like these other examples of reduction, whether probability is thus
explained, or explained away, it can no longer be viewed as fundamental. It
can only have the status of the branching structure itself; it is `emergent'. Chance, like quasiclassicality, is then an
`effective' concept, its meaning at the microscopic level entirely
derivative on the establishment of correlations, natural or man-made, with
macroscopic branching. That doesn't mean amplitudes in general (and other
relations in the Hilbert space norm) have no place in the foundations of EQM
-- on the contrary, they are part of the fundamental ontology -- but their
link to probability is indirect. It is simply a mistake, if this reduction
is successful, to see quantum theory as at bottom a theory of probability.\

\section{Explaining probability}

Functional reduction is not new to the sciences; functional reduction,
specifically, of probability, is not new to philosophy. In any moderately
serious form of physical realism a world is a system of actual events,
arranged in a spatiotemporal array, defined in terms of objective, physical
properties and relations alone. Where in all this are the physical
probabilities? For physicalists, they can only be grounded on the actual
structure of events - what Lewis has called a `Humean tapestry' of events
(Lewis [1986a xv-xvi]), whereupon the same identificatory
project ensues as in EQM\footnote{%
Lewis explicitly contemplated extending his `tapestry of events' to quantum
mechanics, acknowledging that novel quantum properties and relations (like
amplitudes and relations among amplitudes) may have to be included; but only
come the day that quantum mechanics is `purified' (Lewis [1986a p.xi]) --
`of instrumental frivolity, of doublethinking deviant logic, and -- most of
all -- of supernatural tales about the power of the observant mind to make
things jump'. Quantum mechanics has certainly been purified in EQM.}. The
links (i), (ii), (iii) stand in need of explanation whatever the physics,
classical or quantum, one world or many. For the most part they have
remained surprisingly hard to explain, even, and perhaps especially, in
one-world theories.\footnote{%
According to even so committed an empiricist as B. van Fraassen, a model of
a probabilistic physical theory must include elements representing
alternative possible sequences of outcomes, and the theory can be true `only
if alternative possible courses of events are real.' (van Fraassen [1980
p.197]).}

But \textit{one} of the three links, in the case of one-world theories, 
\textit{is} easily explained -- namely (iii), the link with uncertainty.
Only allow that the dynamics is indeterministic (whether in a fundamental or
effective sense), so that some kinds of events in the future are (at least
effectively) unpredictable, then, for the denizens of that world, the future
will in some respects at least be uncertain. This is how all the
conventional theories of chance explain (iii) (with varying strengths of
`unpredictability'). But according to EQM all outcomes exist, and, we may as
well suppose, are known to exist (we may suppose we know the relevant
branching structure of the wave-function) -- so the future is completely
known. There is no uncertainty on this score.

This marks out EQM\ as facing a particular difficulty with (iii), the link
with uncertainty, a difficulty encountered by none of the familiar theories
of probability in physics. It has seemed so severe, in fact, that not only
do the majority of critics of EQM believe that as a theory of probability it
must rest on the links (i) with statistics and (ii) rationality alone (if it
is to mean anything), but so do many of its friends.\footnote{%
E.g. Papineau, [1996, 2010], Greaves [2004], [2007], Deutsch [1999, 2010].} The
attempt to ground EQM on (i) and (ii) alone, disavowing all talk of
probability or uncertainty, has been dubbed the \textit{fission programme}.
This article is intended as a counterbalance, and is skewed towards a defense of (iii), the
link with uncertainty. As I shall argue in Section 3 and 4, the status of
this link is not in fact so different in EQM than in conventional theories
of chance, using a possible-worlds analysis, of a sort familiar to
philosophers.

I have less to say on (i) and (ii), not because they are less important to
the overall argument, but because arguments for them are either well-known
(in the case of (i), the link with statistics) or given elsewhere in this
volume (by Wallace, for (ii), the principal principle). Both links now
appear to be in significantly better shape in EQM\ than in the other main
theories of physical probability (classical statistical mechanics,
deterministic hidden variable theories, and stochastic physical theories),
none of which offer comparably detailed explanations of (i) and (ii).
Instead they presuppose them.

First (i), the link with statistics. A first step was the demonstration that
a largely formal condition of adequacy could be met: a quantum version of
the Bernouilli (`law of large numbers') theorem could be derived\footnote{%
As first sketched by Everett in the `long dissertation' (Everett [1973]).}. This says that in the limit of large numbers of trials, the probability of relative  frequencies different from the predicted probabilities goes to zero. But this theorem follows from the rules of probability. Distinctive, in EQM, is that given an account of what chance processes /textit{actually are} (branchings), and given that in EQM we can model any measurement process as comprehensively as we please (including `the
observer' if need be) it becomes a purely\textit{\ dynamical} question as to
whether and how branch amplitudes can be measured. I will give some
illustrative examples in Section 2 -- but I take these arguments to be
relatively uncontroversial.

Not so the arguments for (ii), the decision theory link. But even here, few
would deny that there has been progress -- and on two quite separate fronts,
both involving the principal principle: the first (the sense already
suggested) in showing why credence should track chance, as identified in our
Everettian tapestry; the second in showing how we could have been led, by
rational, empirical methods, to anything like that tapestry in the first
place.

For the former argument, see Wallace [2010, 2012], deriving the principal
principle.\footnote{%
Note added Sep 2016. This is the principal principle in the special instance of EQM. Wallace speaks instead of `deriving the Born rule', but of couse there are many derivations of the Born rule (that have nothing to do with decision theory): of which Everett [1957] was the first. See e.g. Zurek [2005], for a derivation from locality, or Saunders [2004, 2005], for a derivation from operational and invariance principles respectively; Another notable [and greatly neglected] derivation was given by Lubkin [1979] (my thanks to Ted Jacobson for pointing this out to me.} He shows that branching structures and the squared moduli of the
amplitudes, insofar as they are known, \textit{ought} to play the same
decision theory role (ii) that chances play, insofar as they are known, in
one-world theories. Whatever one might think of Wallace's axioms, as long as
nothing underhand or sneaky is going on the result is already a milestone:\
nothing comparable has been achieved for any other physical theory of chance.

It is, however, limited to the context of agents who believe the world has
the branching structure EQM says it has --\ it is a solution to the
`practical problem', the normative problem of how agents who believe they
live in a branching universe ought to achieve their ends (without any prior
assumption about probability). It has no direct role in confirming or
disconfirming EQM as one of a number of rival physical theories by agents
uncommitted as to its truth. It leaves the `evidential problem' unsolved.

But on this front too there has been progress. Greaves [2007] and
Greaves and Myrvold [2010] showed how confirmation theory (specifically
Bayesian confirmation theory) can be generalized so as to apply to branching
theories and non-branching theories evenhandedly, without empirical
prejudice. The latter proved a representation theorem on the basis of axioms
that are entirely independent of quantum mechanics. Nor, in keeping with the
fission programme, do they make explicit or even tacit appeal to the notion
of probability or uncertainty.

Wallace's axioms were likewise intended to apply even in the context of the
fission programme. If acceptable that shows the argument for (ii) is independent of (iii).
But if the argument for (iii) (with (i)\ and (ii)) goes through, then the
present approach \textit{also} promises a solution to the evidential
problem. As an objection to the Everett interpretation, the problem only
arises if it is granted that branching and branch amplitudes cannot be
identified with chance and probabilities (which we take to mean are not in
fact quantities that satisfy (i), (ii), (iii)), precisely the point here in
contention. Show that they can, and the evidential problem simply dissolves.
Or rather, since it can be read as a problem for every chance theory, it
will have the same status in the Everett interpretation as it has in any
other physical theory of chance, to all of which the Greaves and Myrvold
analysis applies.

I have little more to say about (ii), the decision-theory link. What follows
is directed to (i), the link with statistics (Section 2), but mainly to
(iii), the link with uncertainty.\footnote{%
Note added Sep 2016. Of Wallace's axioms, `branching indifference' (that branching per se, where the differences between branches do not matter, should be treated as irrelevant) has been thought particularly vulnerable. But if uncertainty is the right attitude in the face of branching, then I \textit{should} be indifferent as to which of a number of persons is me, if they differ only in ways that do not matter to me. See Wilson [2011], [2013].}

 In Section 3 I show how branching in EQM
is consistent with talk of uncertainty (following Saunders and Wallace
[2008a]). That is arguably enough, with the results of Section 2 and in decision theory, 
 to draw our main conclusion: branching in EQM should
be identified with chancing, and mod-squared branch amplitudes with chance.

But there remains a contrary reading, according to which there is no place
for uncertainty after all -- or not of the right kind. Given sufficient
knowledge of the wave-function, on this contrary reading uncertainty can at
best concern \textit{present} states of affairs, not future ones. In the
final and more philosophical section I\ argue that the difference boils down
to a choice of metaphysics, a choice that is under-determined by the
mathematical structure of the theory. Since choices like this should be made
to help in the understanding of a physical theory, rather than to frustrate
it, the contrary reading should be rejected.

\section{Why chance is measured by statistics}

Here \noindent are a number of no-go facts about how physical probabilities
are observed that we believe to be true, but that seem very difficult to
explain on any conventional theory of physical probability:

\begin{enumerate}
\item[(a)] There is no probability meter that can measure single-case chance
with chance equal to one.

\item[(b)] There is no probability meter that can measure chance on repeated
trials with chance equal to one.

\item[(c)] There is no probability meter that can measure single-case chance
with chance \textit{close} to one.

\item[(d)] Absolute probability can never be measured, only relative
probability.
\end{enumerate}

\noindent (The list could be continued.)

On conventional thinking, someone who needs convincing of these facts about
chance has not so much failed to understood a physical theory as the \textit{%
concept} of probability. But if chance is a definite physical magnitude, it
should (at least indirectly) be measurable, like anything else. Given the
dynamics, and a theory complete enough to model the experimental process
itself, these facts should be explained. And indeed no non-trivial function
of the branch amplitudes can be measured in the ways mentioned in (a)-(d),
according to EQM. This is a \textit{dynamical} claim.

On the positive side, as to how (we know) chances are in practice measured:

\begin{enumerate}
\item[(e)] There are probability meters that measure chances on repeated
trials (perhaps simultaneous trials) with chance close to one.
\end{enumerate}

\noindent \noindent This is conventionally thought of as a consequence of
the axioms of probability (the law of large numbers, or Bernouilli theorem)
rather than of any physical theory; in turning it around, and deriving it
rather from the dynamical equations of EQM, it identifies the appropriate
physical quantities that are to count as probabilities instead.

There remains one other obvious fact without which one might well think no
account of chance can so much as get off the ground: chance outcomes are
typically \textit{mutually exclusive} or \textit{incompatible} -- in which
case only \textit{one} outcome obtains (which one happens by chance). That
is (where I explicitly add the qualification `observable', as we are
concerned with how chances can be measured):

\begin{quote}
\textbf{Presupposition}: Of two incompatible, observable, chance events,
only \textit{one} event happens
\end{quote}

How can the presupposition be explained in EQM? In comparison with (a)-(e)
it seems to have the least to do with chance -- the word `chance' could be
deleted from it entirely -- but we surely do talk about chance in this way.
It is presupposed by any application of the concept of chance that (at least
sometimes) chances apply to incompatible outcomes. The concept of
incompatibility enters at the the very beginning of any mathematical
definition of a probability measure on a space of events (events are
represented by sets, and inherit from set theory the structure of a Boolean
algebra). And yet it seems to be violated straightforwardly in EQM, as
Everett's crazy idea was that in a quantum measurement \textit{all} outcomes
happen.

If that was all there was to it it would be hard to understand why EQM was
ever taken seriously by anyone. The answer is that the presupposition has
two different readings, the one a physical or metaphysical claim -- about
what chances fundamentally are -- and the other a claim about what is
observable (by any observer), the phenomenology of chance.

On the first reading, the presupposition \textit{is} straightforwardly
denied; it is denied if the word `chance' is deleted too. But we already
knew this: this is simply a conflict between a many-worlds and a one-world
theory. Likewise for the presupposition as a metaphysical claim:\ we already
knew EQM challenges a number of a priori claims. It is only as an
epistemological claim -- as to the \textit{observed} phenomenology -- that
the presupposition had better still make sense. But so it does: EQM explains
it very simply. No two incompatible chance outcomes can happen \textit{in
the same branch}. And since the apparatus, and the observer, and the entire
observable universe, are branch-relative -- they are `in-branch' structures
-- no observer can ever witness incompatible outcomes simultaneously.%
\footnote{%
If this is thought to be question-begging: there does not exist a perception
of two incompatible outcomes, according to EQM, treating perception
biochemically, or indeed as any kind of record-making process. (This point
goes back to Everett [1957] and, ultimately, the von Neumann model of
measurement. See also Gell-Mann and
Hartle [1990] and Saunders [1994].)} In this second sense, then, the
presupposition is rather elegantly dealt with in EQM -- `elegantly', because
it follows directly from its account of what a chance set-up is, and from
what chance is (one should \textit{not} delete the word `chance'!).

The other everyday (`no-go') facts about probabilities follow, not from
branching alone, but from the unitarity of the equations of motion. Because
I do not think this claim is particularly controversial I will simply
illustrate it with a proof of (a), for relative probabilities (meaning a quantity that concerns the relationships between chancy outcomes).Consider then
a microscopic system in the state 
\begin{equation*}
|\varphi _{c}\rangle =c|\varphi _{+}\rangle +\sqrt{1-|c|^{2}}|\varphi
_{-}\rangle .
\end{equation*}%
For simplicity, model the measurement process as a 1-step history, using the
Schr\"{o}dinger picture, with initial state $|\omega \rangle \otimes
|\varphi _{c}\rangle $ at $t=0$ where $|\omega \rangle $ is the state of the
rest of the universe. Let the configuration describing (among other things) the apparatus registering (a function of) the amplitude $c$ at time $t$ be $%
\alpha (c)$. Let the projection onto this configuration be $P_{\alpha (c)}$
and let the associated Heisenberg-picture projection be 
\begin{equation*}
P_{\alpha (c)}(t)=e^{iHt/\hbar }P_{\alpha (c)}e^{-iHt/\hbar }
\end{equation*}%
(See appendix 1 for how quantum histories are defined as time-ordered products of such projectors.) The no-go fact we need to establish is that there is no unitary dynamics $%
U_{t}$ $=e^{-iHt/\hbar }$ by means of which:

\begin{equation*}
|\omega \rangle \otimes |\varphi _{c}\rangle \rightarrow U_{t}|\omega
\rangle \otimes |\varphi _{c}\rangle =U_{t}P_{\alpha (c)}(t)|\omega \rangle
\otimes |\varphi _{c}\rangle
\end{equation*}%
for variable $c.$ If there were it would also follow that for $c^{\prime
}\neq c$:%
\begin{equation*}
|\omega \rangle \otimes |\varphi _{c^{\prime }}\rangle \rightarrow
U_{t}P_{\alpha (c^{\prime })}(t)|\omega \rangle \otimes |\varphi _{c^{\prime
}}\rangle .
\end{equation*}%
But the inner product of the LHS of the two initial states can be as close
to one as desired, whilst that of the vectors on the RHS must be zero, as $%
\alpha (c)$ and $\alpha (c^{\prime })$ must differ macroscopically for
sufficiently large $|c-c^{\prime }|$, a contradiction. Thus not even
relative probabilities can be measured deterministically. (The argument for
absolute probabilities is even simpler, and depends only on the linearity of
the Schr\"{o}dinger equation.)

Now for (e), how (relative) chances can be measured according to EQM, with high chance.
Take a number $N$ of subsystems, each (to a good approximation) in state $%
\varphi _{c}=c\varphi _{+}+\sqrt{1-|c|^{2}}\varphi _{-}$, and arrange the
linear dynamics so that%
\begin{equation}
|\omega \otimes \varphi _{\pm }\rangle \rightarrow U_{t}P_{\alpha (\pm
)}(t)|\omega \otimes \varphi _{\pm }\rangle
\end{equation}%
i.e. the von Neumann model of measurement, where $\alpha (+)$ is a
configuration in which the apparatus reads `spin up' and $\alpha (-)$ `spin
down'. Applied to an initial state of the form $|\omega
\rangle \otimes |\varphi _{c}\rangle $, it yields the superposition 
\begin{equation*}
|\omega \rangle \otimes |\varphi _{c}\rangle \rightarrow U_{t}P_{\alpha
(+)}(t)|\omega \otimes \varphi _{c}\rangle +U_{t}P_{\alpha (-)}(t)|\omega
\otimes \varphi _{c}\rangle
\end{equation*}%
in which the first vector has norm $|c|$ and the second has
norm $\sqrt{1-|c|^{2}}.$ That is, the amplitudes $c$, $\sqrt{1-|c|^{2}}$ of
components in a microscopic superposition have been promoted to macroscopic 
\textit{branch} amplitudes. Consider now repeated measurements of $N$
microscopic systems all in the same state $|\varphi _{c}\rangle $, whether
sequentially repeated in time, or measured all at once. The latter is the
simplest to model, assuming the $N$ apparatuses are non-interacting: the
result at time $t$ will be a superposition of vectors at $t$ of the form
\begin{equation*}
|\alpha _{f}(t)\rangle =U_{t}P_{f(1)}(t)\otimes ...\otimes P_{f(N)}(t)|\omega
\rangle \otimes |\varphi _{c}\rangle \otimes ...\otimes |\varphi _{c}\rangle
\end{equation*}%
where $f(k)$, $k=1,...,N$ is either +1 or -1. Those with the same relative
frequencies $M/N$ will all have the same norm, i.e.:%
\begin{equation}
\left\vert |\alpha _{f},t\rangle \right\vert =|c|^{M}\sqrt{(1-|c|^{2})}%
^{N-M};\text{ }\sum_{k=1}^{N}f(k)=2M-N.
\end{equation}%
The unitary evolution to time $t$ is: 
\begin{equation}
|\omega \rangle \otimes |\varphi _{c}\rangle \otimes ...\otimes |\varphi
_{c}\rangle \rightarrow \sum_{M=0}^{N}\left[ \sum_{f;\text{ }\Sigma
_{k=1}^{N}f(k)=2M-N}|\alpha _{f},t\rangle \right] .
\end{equation}%
The right-most summation, for fixed $M,N$, is over all $N!/M!(N-M)!$
distinct $f$'s, all with the same norm Eq.(2). Since $\langle \alpha _{f},t$ 
$|\alpha _{g},t\rangle =0$ for $f\neq g$,\footnote{%
For sequences of measurements in time, the consistency condition is needed
at this point: see appendix 1.} the squared norm of the RHS of Eq.(3) is

\begin{equation*}
\sum_{M=0}^{N}\frac{N!}{M!(N-M)!}\left\vert |\alpha _{f},t\rangle
\right\vert ^{2}=\sum_{M=0}^{N}x(M)
\end{equation*}%
where $x(M)$ is:%
\begin{equation*}
x(M)=|c|^{2M}(1-|c|^{2})^{N-M}\frac{N!}{M!(N-M)!}.
\end{equation*}%
For $N$ large, this function is strongly peaked about $M/N=\left\vert
c\right\vert ^{2}.$ The sum of norms of vectors with the `right' relative
frequency is much larger than the sum of norms of vectors with the `wrong'
relative frequency. In this sense measured relative frequencies close to $%
\left\vert c\right\vert ^{2}$ are found on relatively high amplitude
branches.

It was rather crucial to this argument that we consider only relative
frequency -- we don't care about precisely which particles were recorded as
`spin-up' and `spin-down' (or which of our $N-$apparatuses recorded which
result). That is, our measurement protocol requires that the measurements be
treated as `exchangeable', in de Finetti's sense (Greaves and Myrvold [2010 pp.277-80]). This is an
axiom of de Finetti's approach, whereas in EQM it is explained.

Might some other method be found by which functions of amplitudes may be
measured, flouting (a)-(d)? Perhaps;\footnote{%
For example, `protective measurements' (Aharonov and Vaidman [1993]),
although see the criticism of Uffink [2000]); or `weak measurements'
(Aharonov et al [1988]).} but that is unlikely to make for objection if,
like EQM, it is based on the unitary formalism of quantum mechanics. If
based on a rival theory, which is empirically successful, that will anyway
spell the end of quantum mechanics (and with it EQM).

Now suppose, fancifully if you will, that we are entitled to identify
chances with functions of branch amplitudes. Then in summary we have just
shown: chances will generally be associated with incompatible observable
outcomes, only one of which can happen at each time; they may not be
measured in a way that is not itself chancy, and single-case chance cannot
be measured at all. They can only be measured by running a chance
(branching) process repeatedly, or by a single trial involving a large
number of similarly prepared systems, and listing the relative frequencies
(taking care to neglect the order of outcomes, or which system had which
outcome). The measurement will be veridical, however, only with high chance,
given the number of systems involved is sufficiently large. None of these
facts can be explained by any conventional physical theory of probability
(rather, they are presupposed).

\section{Why chance involves uncertainty}

We cannot, however, identify chance set-ups with branch set-ups, and chances
with functions of amplitudes, failing an account of (iii), the link with
uncertainty. In the absence of this, to arrive, by the Deutsch-Wallace
representation theorem, at an agent's credence function (in conformity with
the Born rule), raises a puzzle on its own. For what does a rational agent
believe to some degree when she uses that credence function? Given the
theorem, certainly we can explain credence by reference to behaviour; but in
this case, better, perhaps, to simply explain it away. A rational agent must
order her priorities somehow, whether or not there is anything of which she
is uncertain; but it is hard to see how is it possible to have degrees of
belief different from one and zero if you know everything there is to know.

The functionalists' response is to more or less recapitulate the
representation theorem: credences derive their meaning from their function
in rational action, in determining expected utilities, and hence agents'
preferences. They will also note that meaning is determined by use -- that
if anything has been agreed by philosophers of language in recent decades it
is:

\begin{quotation}
\noindent Words have no function save as they play a role in sentences:
their semantic features are abstracted from the semantic features of
sentences, just as the semantic features of sentences are abstracted from
their part in helping people achieve goals or realize intentions. (Davidson
[2001, p. 220])
\end{quotation}

\noindent So there is every reason to talk of expected utilities and
credences. By the same broadly functionalist philosophy, if these credences
play the same role that credences about chance events play, they \textit{are}
credences about chance events. Talk of uncertainty and unpredictability then
falls into place along with the rest of our ordinary use of words.

I think this argument is essentially correct, but it leaves unanswered --
brief explanations please! -- just \textit{how} talk of uncertainty is to
fall into place, and just \textit{what} to say in answer to the question of
what these degrees of belief are about.

First a warm-up. In the previous section I argued that functions of branch
amplitudes are measured -- they are manifested in-branch -- in just the way
that chances are measured. A part of that argument was that the
presupposition about incompatible outcomes follows (rather
straightforwardly) from EQM: on each trial, since macroscopically distinct,
only one outcome can be obtained in each branch. So from the point of view
of accumulating information that can actually be used -- `in-branch'
information -- events are registered sequentially, just as they are in a
non-branching theory. This was what Everett [1957] took pains to show. But
isn't sequential increase in information increase in knowledge? So isn't
there something that is being learned in each branch?

Consider a concrete example. Alice, we suppose, is about to perform a
Stern-Gerlach experiment. She understands the structure of the apparatus and
the state preparation device, and she is convinced EQM\ is true. In what
sense does she learn, post-branching, something new?\ The answer is that 
\textit{each} Alice\textit{, }post-branching\textit{, }learns something new
(or is in a position to learn something new) -- each will say something
(namely, `I see the outcome is spin-up (respectively, spin-down), and not
spin-down (respectively, spin-up)') that Alice prior to branching cannot
say. It is true that Alice, prior to branching knows\textit{\ that} this is
what each successor will say -- but still she herself cannot speak in this
way.

To make this vivid, imagine that each of her successors simply 
\textit{closes her eyes}, when the result is obtained; in that state, each
is genuinely ignorant of the result (so-called `Vaidman ignorance'; Vaidman
[1998]). This has usually been considered (by Vaidman himself) as ignorance
that Alice, prior to branching, does not have, but it can be turned around
the other way: each Alice, closing her eyes, perpetuates a state of
ignorance that she already had.

The implication of this line of thought\footnote{%
Picked up also by Ismael [2003], although she develops it in a rather
different way.} is that appearances notwithstanding, prior to branching
Alice does \textit{not} know everything there is to know. What is it she
does not know? I say `appearances notwithstanding', for of course in one
sense (we may suppose) Alice does know everything there is to know;\ she
knows (we might as well assume) the entire corpus of impersonal, scientific
knowledge. But what that does not tell her is \textit{just which person she
is} -- or \textit{where she is located} \ -- in the wave-function of the
universe.

The point is a familiar one to philosophers. One can know that there are
such-and-such people, but not which one of them is me. Or that such-and-such
events occur at various places, but not which of those places is here; or
times, but not which of those times is now.

Such knowledge that is omitted is sometimes called `indexical' knowledge, by
philosophers, also `knowledge de se', and `self-locating knowledge' (we
shall use the latter). But why should knowledge like this be lacking?\
Vaidman has Alice hide her eyes. Another example, due to J. R. Perry,
likewise suggests some kind of impairment is needed. Perry asks us to
consider an amnesiac who has lost his way in the library at Stanford. He
does not know who he is; he does not know where he is -- not even were he to
read every book in the library, not even if he were to read his own \textit{%
biography}, would he be any the wiser. But Alice is not so impaired. She
does not hide her eyes, and no more is she an amnesiac. She sees where she
is in the wave-function of the universe and self-locates accordingly -- and
differently -- from where her successors locate themselves, for, obviously,
she is at a different place from them. She is not ignorant of anything her
successors know; she simply reports different self-locating facts from them.

And so she would, \textit{but only given certain assumptions as to what,
exactly, she is -- of how she is represented in the physics}. For example,
if she at $t_{j}$ is represented by the configuration $\beta _{j}$, or by
the Heisenberg-picture vector $|\beta _{j}\rangle =P_{\beta
_{j}}(t_{j})|\Omega \rangle ,$ where $|\Omega \rangle $ is the universal
state, with her successors at $t_{k}>t_{j}$ similarly represented by
configurations $\beta _{k}$, $\beta _{k}^{\prime }$, or vectors $|\beta
_{k}\rangle ,|\beta _{k}^{\prime }\rangle $, the result surely follows. But
this treats her as strictly a momentary thing, independent of the history in
which she is located. There is an alternative: if she at $t_{j}$ is
represented instead by an entire history $\beta $, indexed by $t_{j}$, then of course, at $t_{j}$, she does not know
which history or branch vector is hers, and quite the opposite result
follows.

Which of the two is correct?\ But questions like these (or rather their
classical counterparts) are well known in metaphysics (and specifically
personal identity). Most philosophers are agreed they cannot be settled on
the basis of the physics alone. Here then is a metaphysics friendly to the
Everett interpretation: let persons be spacetime worldtubes from cradle to
grave (in the jargon, `maximal continuants'). If Alice is a person (of
course she is) then we must say, even prior to branching at $t_{j}$, that
there are \textit{many} Alice's present, atom-for-atom duplicates up to $%
t_{j}$, each behaving in exactly the same way and saying just the same
words. If that is the right metaphysical picture for EQM, then Alice \textit{%
should }be uncertain after all -- \ \textit{each} Alice should be uncertain
-- for each as of $t_{j}$ does not (and as a matter of principle, cannot)
know which of these branching persons is she.

That, I take it, establishes that the question of whether or not there is
uncertainty in the Everett interpretation can be settled either way,
depending on a metaphysics of personal identity. But rather than invoke a
metaphysical assumption, we can make do with a different kind of claim -- a
proposal, not about the ultimate natures of persons, but about the reference
of the word `person' in EQM terms. The proposal is that talk of persons (and things)
be relativized as to branch:\footnote{%
This semantics was discussed briefly in Wallace [2005], [2006] and in more
detail in Saunders and Wallace [2008a], on which the argument that follows
is based.}

\begin{description}
\item[S1] By a 'person' or `thing' is meant a branch-part or ordered pair $%
(\beta $,$|\alpha \rangle )$, where $\alpha \in \beta .$
\end{description}

\noindent Here `$\alpha \in \beta $' means the sequence of configurations $%
\beta $ is obtained by a coarse-graining of $\alpha $, which we suppose is
temporally and spatially much more finely grained. Likewise `a person at
time $t_{j}$' is represented by an ordered triple $(\beta
_{j},\beta ,|\alpha \rangle )$ (if a person is an entire history), where $\alpha \in \beta \in \beta _{j}$, or by an ordered pair $(\beta _{j}$,$|\alpha \rangle )$ (if a person is a stage). More is needed for
tensed sentences (involving `was', `is', `will be' etc.), but for this, and
on the question of how S1 can be justified, see Section 4.

For now note $S1$'s virtues. First, it is clearly permissible; it makes use
of nothing but the available mathematics of EQM in terms of branch vectors $%
|\alpha \rangle $ and sequences of configurations $\alpha $, $\beta $ etc.,
as ordered pairs $\{(\beta ,|\alpha \rangle );$ $\alpha \in \beta \}.$%
\footnote{%
Contrary to Kent [2010 p.346].}. True, this requires the
consistent histories formalism, in which branch vectors are
Heisenberg-picture vectors (so that a branch vector $|\alpha \rangle $
describes the entire history $\alpha $, not any particular instant of it).
The perspective is atemporal. But branching itself is a process defined by
the dynamics over time: the idea of a decoherence basis, defined at an
instant of time independent of what comes before and what comes after, is a
fiction.

Second, using $S1$, not only is Alice entitled -- bound -- to be unsure of
the outcome of the experiment, but she has a genuine gain in knowledge when
it is learned. For after branching, using S1, on observing the outcome, each
Alice self-locates better than she did before -- knows more than she did
before -- and so has learned something that, prior to branching, she could
not have known. Vaidman's ignorance is ignorance each Alice already had.
Likewise, there can be no algorithm, whose input is data about a branch up
to one time, and whose output uniquely specifies that branch at a later
time; for the same algorithm must operate in every other branch which is
exactly the same up to that time but which differs thereafter. Branching
events are algorithmically uncertain too -- they are \textit{indeterministic}%
.

Third, $S1$ is neutral on some (but not all) metaphysical questions about
personal identity. Specifically, those that arise given $4-$dimensionalism,
assuming a single world, are likely to play out the same under $S1$, in the
context of branching worlds, if only it is permitted to relativize one's
favoured candidate for $\beta $ to worlds (represented by Heisenberg-picture
branch vectors $|\alpha \rangle $, where $\alpha \in \beta $).

And the bottom line: under $S1$, prior to branching, uncertainty is assured:
Alice doesn't know if she will see spin-down or spin-up, as she doesn't know
which branch she is in. Unless some hidden contradiction is involved, $S1$
is the right rule for making sense of quantum mechanics in realist terms.

\medskip

\noindent \textbf{Conclusion. }Branching, the development of superpositions
of the universal state with respect to the decoherence basis, plays all the
chance roles (i), (ii), (iii): it produces the same phenomenology as chance, the squared norms of branch amplitudes are of practical relevance to decision
theory in the way that chances normally are, and using SI, 
branching involves uncertainty, just as chances do. In
EQM, branching and squared norms of branch amplitudes are demonstrably
functionally equivalent to chance, in these three central respects;
therefore they \textit{are} chance processes, and chances \textit{are}
these physical magnitudes.

\section{Overlap and divergence}

\noindent There are alternatives to the rule $S1$, however. Invoking it
seems to compromise a chief selling point of the Everett interpretation,
which is that many worlds follows from the unitary dynamics, with no added
principles or special assumptions. That is what puts the Everett
interpretation\ in a class of its own when it comes to the quantum realism
problem: there are plenty of avenues for obtaining (at least
non-relativistic) one-world theories if we are prepared to violate this
precept.

On the other hand $S1$ is on the face of it \textit{just} a semantic rule --
it is \textit{merely} a linguistic matter. The referents of terms are
constrained by their contexts of use, granted; but over and above those
constraints, not even in the God's eye view is their meaning determined. It
is up to us to say what, precisely, among the structures in the universal
state, our words really mean. And if that is all there is to uncertainty,
nothing much should hang on the matter either way. The challenge remains: to
justify the probability interpretation of EQM on the grounds of (i) and (ii)
alone.

I think there is something right about this argument. Whether or not there
is genuine uncertainty in the Everett interpretation appears rather less
substantive than might have been thought. But in that case, let satisfaction
of the chance roles (i) and (ii) be enough to count for branching to count
as chance; our conclusion stands.\footnote{%
This is not the fission programme, which would have us renounce talk of
objective chance and genuine probability (along with uncertainty).}

But on three counts this would be too quick. The first is that for many, $S1$
\textit{really is} a metaphysical claim, for all my talk of semantics -- and
that there \textit{is} a substantive question of whether it is true.
Certainly there are alternative metaphysical claims that could be made that
cause problems to $S1$. I shall have something to say about this sort of
argument, but not much: for I do not believe there are metaphysical truths
of this sort, independent of ordinary language and natural science. Or let
me put the point more constructively: metaphysics should primarily be
answerable to language and to science; it should be `naturalized' (Saunders
[1997], Ladyman and Ross [2007]). If this is right $S1$ may still be a
metaphysical claim, a claim of substance, but in the naturalized sense.

The second count is that we should take rather more seriously the task of
accounting for language use, for that too is emergent structure, a part of
the physical world to be studied as such by scientific means. As with any
form of functional reduction, we know what we are looking for in advance:
the problem is to identify the right sort of structure at the more
fundamental physical level to account for phenomena that we are already
familiar with at the less. Devising semantic rules, whose truth conditions
are fixed by reference to the underlying theory, is only a variation on the
same procedure -- it is reductionism as it is appropriate to linguistics. It
is no different in kind from fixing on certain variables, for example
hydrodynamical variables, in decoherence theory, to derive
quasiclassicality. Sure, there remain the \textit{wrong} variables, ones
that give no hint of classicality; and wrong semantic rules too, which are
inadequate to explaining our linguistic behaviour -- and which give no hint
of uncertainty. They should be eschewed.

This argument is a variant of Wallace's [2005], which puts the matter in
terms of `the principle of charity'. This principle says that the most
important criterion of `good' translation, in the radical case, where no
prior standard of translation has been established, is that it maximizes
truth. Working out what to say about branching, if EQM is true, is like
radical translation -- of how to translate into our own tongue the
expressions of some alien language on the basis of observable linguistic
behaviour alone. If EQM is true, it has always been true, and we have always
used words like `uncertainty' and `chance' in the context of branching so --
like it or not! -- that is what those words have always been about. Whatever
we might say about novel scenarios like teleportation machines or brain
transplants (the ways in which philosophers have tended to imagine persons
being divided) our use of words in ordinary contexts is not in doubt. The
referents of ordinary words and truth conditions of everyday sentences (as
specified in a physical theory) may wait to be correctly identified, but the
criteria for `correctness' is decided, not by metaphysics, but by their
adequacy, as follows from the theory, to make sense of what we ordinarily
say.

Greaves herself accepts this argument (Greaves [2007 p.124]), but not all
Everettians do, and surely not all skeptics of EQM\ do. But this dispute is
independent of EQM per se. It should be settled if possible by some other
example of the general method -- say, in the arena of a semantics for
temporal affairs, involving tense and becoming, in classical spacetime
theories (as argued by Saunders [1996], Wallace [2005]), or in the arena of
a semantics for agency, moral responsibility, and free-will, in
deterministic theories.

The third count on which the case for uncertainty can be too quickly
deflated is a strengthened version of the first. I have already hinted at
it:\ maybe $S1$ \textit{does} harbour some hidden contradiction, or some
failure of integrity more broadly construed. The semantics should not 
\textit{misrepresent} our situation, were EQM true.

This, it seems to me, is the only serious concern, and the only one I shall
consider in the rest of this chapter. But it needs explaining. The very
worry seems strange. How can a sentence misrepresent a theory, if it is true
by that very theory?

Some more detail on the semantics will be helpful\footnote{%
Other rules may be possible as well: see Wallace [2006] for some
alternatives. The rules that follow are illustrative.}. Let $\alpha =\langle
\alpha _{+},..,\alpha _{-}\rangle $, where $t_{+}$ is much later than any
time we are interested in, and $t_{-}$ is much earlier. Call branches like
this \textit{maximal}. By `world' I mean maximal branches, represented by
maximal branch vectors. Let $F$ be true of some branches, false of others.
Then since, by $S1$, speakers and things are parts of branches, so are
utterances:

\begin{description}
\item[S2] An utterance of `$F$' in branch $|\alpha \rangle $ is true if and
only if $F$ is true in $\alpha .$
\end{description}

\noindent Now for tensed statements. Let $\alpha ^{+}(t)\underset{def}{=}%
\langle \alpha _{+},..,\alpha _{k}\rangle \ni \alpha $, $t_{k}>t\geq t_{k-1}$
be \textit{the future of} $\alpha $ \textit{at} $t$; let $\alpha ^{-}(t)%
\underset{def}{=}\langle \alpha _{k},..,\alpha _{-}\rangle \ni \alpha $, $%
t_{k+1}>t\geq t_{k}$ be the \textit{past of }$\alpha $ \textit{at} $t$.
Let $F$ be true at some times, false at others (it is an
`occasion sentence', in Quinean terms). The rule for future physical
contingencies is:

\begin{description}
\item[S3] An utterance of `$F$ will be the case' in branch $|\alpha \rangle $
at $t$ is true if and only if $F$ is true in the future of $\alpha $ at $t$.
\end{description}

\noindent A first stab at a rule for future possibilities is:

\begin{description}
\item[S4] An utterance of `$F$ might happen' in branch $|\alpha \rangle $ at 
$t$ is true if and only if for some branch $|\alpha ^{\prime }\rangle ,$ $F$
is true in the future of $\alpha ^{\prime }$ at $t$, where the past of $%
\alpha ^{\prime }$ and $\alpha $ at $t$ is the same.
\end{description}

\noindent But the latter will obviously need to be restricted to branches
vectors $|\alpha ^{\prime }\rangle $ whose norm, conditional on $\alpha
^{-}(t)$, is non-negligible (or else it will turn out that pretty well
anything might happen). And similarly for counterfactuals.

$S2$ and $S3$, fairly obviously, satisfy the principle of charity. $S4$
promises to, at least if `might happen' is taken to mean `might happen by
chance' -- that is, in accordance with the laws of
quantum mechanics. Apart from this proviso, it fits well enough with
standard ideas in modal metaphysics. One has only to replace `branch $%
|\alpha \rangle $' by `possible world' and the `sameness of the past'
relation entering in $S4$ by a `nearest counterpart' relation (Lewis [1973]) -- although philosophers typically are interested in a notion of possibility of much broader scope. 

Now notice that the `sameness of the past' relation is neutral on the
question of whether, in the past, (maximal) branches (or rather
spatiotemporal parts of such branches)\ are \textit{numerically} the same,
or only \textit{qualitatively} the same. They are neutral on the question of
whether (to use a technical term in philosophy) they `overlap' or merely
`diverge'. But this difference is crucial, say metaphysicians. Rules like $S1-S4$ have been offered by philosophers
(almost always) as a semantics for the latter sort, for `diverging' worlds,
not for the former -- or, as philosophers also call overlapping worlds,
`branching' worlds.

Which is it, in EQM?\ This question is in danger of being settled on the
basis of an accident of terminology. Let us agree that `branching' means the
development of superpositions with respect to the decoherence basis. For the
sense intended by philosophers -- where there are numerical identities among
spatiotemporal parts -- we will speak of `overlap'. Mundane examples of
overlap are everywhere. Thus roads overlap if they share the same stretch of
asphalt; houses and roofs, cars and steering wheels, hands and fingers, all
overlap. Overlapping, like Everettian branching, supposedly has a formal
definition too, but not in terms of Hilbert space structure. Rather, it is
defined in terms of `mereology', the general theory of parts and wholes. (I
shall come back to this shortly.)

As for `divergence', its definition is more equivocal. It is sometimes used
to mean that there exist no physical relations between worlds, contrary to
the situation for worlds in EQM\footnote{%
This can cause some confusion, evident in Saunders [1998 Sec.5], where I
erroneously said that `fatalism' (a near-neighbour to the position I am
currently defending) involved the replacement of the superposition of
histories (the universal state) by an incoherent mixture of histories, and
must thus be rejected.}, but we shall take it as simply the opposite of
overlap: `diverging' means `non-overlapping'.

Now to the point:\ if worlds overlap in EQM then (definitionally) there are
genuine transworld identities. The very same thing exists in different
worlds. In particular, since branching is massive in EQM, if worlds overlap
then the very same person is part of vast numbers of worlds, differing,
perhaps, only in respects remote in space and time -- differing, maybe, only
after centuries. All this is contrary to $S1$. In that case any uncertainty
to be had can have nothing to do with a metaphysics of personal identity.
But is there any uncertainty to be had?

The worry is not that overlapping worlds are unintelligible or inconsistent;
it is that they make nonsense of ordinary beliefs. As Lewis put it\footnote{%
Not everyone agrees with Lewis on this point. Thus Johnston [1989] suggested
the rule $S3$ in the explicit context of overlapping persons, arguing that
semantic rules like this were underdetermined by the metaphysics.}:

\begin{quotation}
\noindent Respect for common sense gives us reason to reject any theory that
says that we ourselves are involved in branching [overlapping].... But we
needn't reject the very possibility that a world branches [overlaps]. The
unfortunate inhabitants of such a world, if they think of `the future' as we
do, are of course sorely deceived, and their peculiar circumstances do make
nonsense of how they ordinarily think. But that is their problem; not ours,
as it would be if the worlds generally branched [overlapped] rather than
diverging. (Lewis [1986b p.209].)
\end{quotation}

\noindent Diverging worlds, composed of objects and events that do not
overlap (that are qualitatively but not numerically identical) do not suffer
from this problem.

Of course `common sense' does not cut it much in the physical sciences; and
Lewis' final sentence could not more comprehensively beg the question. But
let us grant this much: he who believes he is contained in each of a number
of worlds cannot also wonder which of them he's in.

We are at the nub of the matter: do worlds -- maximal branches, sequences of
relative configurations of particles and fields, as described by EQM --
overlap in the philosophers' sense, or do they diverge?\ 

We should discount two considerations. First, the coincidence in the
terminology `branching' (because as introduced by Everett, it referred to
the mathematical formalism of quantum mechanics, not to the philosophers'
criterion of overlap). Second, the fact that worlds in EQM do not diverge in
the sense of being physically disconnected (they are not physically
disconnected, because they superpose, but the issue is whether or not they
overlap).

Their remains another consideration, however. There is a clear parallel with
simultaneous rather than temporal overlap. For (ignoring entanglement)\ let
the Schr\"{o}dinger-picture state of a composite system of observer and
environment at time $t$ be
\begin{equation}
|\psi (t)\rangle \otimes (c|\chi (t)\rangle +c^{\prime }|\chi ^{\prime
}(t)\rangle )
\end{equation}%
where $|\psi (t)\rangle $is the state of the observer, and $c|\chi
(t)\rangle +c^{\prime }|\chi ^{\prime }(t)\rangle $ is the state of the
environment, a superposition of macroscopically distinct states $|\chi
(t)\rangle $, $|\chi ^{\prime }(t)\rangle $. Perhaps the latter only differ
with respect to macroscopic objects at enormously large spacelike distances
-- say, a radioactive decay that triggers a macroscopically significant
event on a planetary system in the far side of Andromeda. In such a case, it
is not at all obvious that the observer should be uncertain as to which
branch, $|\psi (t)\rangle \otimes |\chi (t)\rangle $ or $|\psi (t)\rangle
\otimes |\chi ^{\prime }(t)\rangle $, she belongs. Her relative state in
Everett's sense is the superposition $c|\chi (t)\rangle +c^{\prime }|\chi
^{\prime }(t)\rangle $; in this there is no hint of uncertainty.

True enough, but for the mathematical identity:%
\begin{equation*}
|\psi (t)\rangle \otimes (c|\chi (t)\rangle +c^{\prime }|\chi ^{\prime
}(t)\rangle )=c|\psi (t)\rangle \otimes |\chi (t)\rangle +c^{\prime }|\psi
(t)\rangle \otimes |\chi ^{\prime }(t)\rangle .
\end{equation*}%
We can read the relativization the other way: of the observer $|\psi
(t)\rangle $ relative to the environment $|\chi (t)\rangle $ with amplitude $%
c$, and of another observer $|\psi (t)\rangle $ relative to the environment $%
|\chi ^{\prime }(t)\rangle $ with amplitude $c^{\prime }$. It cannot be
necessary to count the two observers as numerically the same, without
further assumptions. Suppose, for example, they have their amplitudes as
properties; if $c\neq c^{\prime }$ they are not even qualitatively the same.

Pursuit of the question of quantum non-locality leads on to relativity,
where it connects with ordinary beliefs about probability in much the same
way that the relativity of simultaneity connects with ordinary beliefs about
tense (Saunders [1995, 1996]). Here we shall stick to probability as it
applies to events related by unambiguously timelike relations, as it figures
in our practical lives.

What is needed is an atemporal perspective. That takes us to the quantum
histories formalism: how does the distinction between overlap and divergence
play out in the Heisenberg picture?

Schr\"{o}dinger picture states at time $t$ are in $1-1$ relation to
histories terminating at $t,$ i.e. of the form $\alpha ^{-}(t)$. We used a
special case of this earlier (for single time histories). More generally, for a history $\alpha=\langle\alpha_{k_n},...,\alpha_{k_1}\rangle$ of configurations at times $t_n,..,t_1$,  let $C_{\alpha}$ be the time-ordered product of the associated Hiesenberg-picture projection operators (a \textit{chain operator}; see appendix 1). The relationship between Schr\"{o}dinger picture states at time $t_n$ resulting from the history $\alpha$ is, up to normalization: 
\begin{equation}
|\alpha ^{-}(t)\rangle =\exp ^{-iHt/\hbar }C_{\alpha ^{-}(t)}|\Omega
\rangle =\exp ^{-iHt/\hbar }\sum_{\alpha \in \alpha ^{-}(t)}C_{\alpha
}|\Omega \rangle .
\end{equation}%
That is, the Schr\"{o}dinger picture state at time $t$ is (the forward
evolution to $t$) of a superposition of Heisenberg-picture states $|\alpha
\rangle =$ $C_{\alpha }|\Omega \rangle $, all with the same sequence of
configurations $\alpha ^{-}(t)$ up to time $t.$ Let `worlds', as before, be
represented by maximal branch vectors $|\alpha \rangle $. Do worlds overlap
or diverge?
\begin{figure}
    \includegraphics[width=12cm, trim=0cm 10cm 2cm 10cm , clip=true]{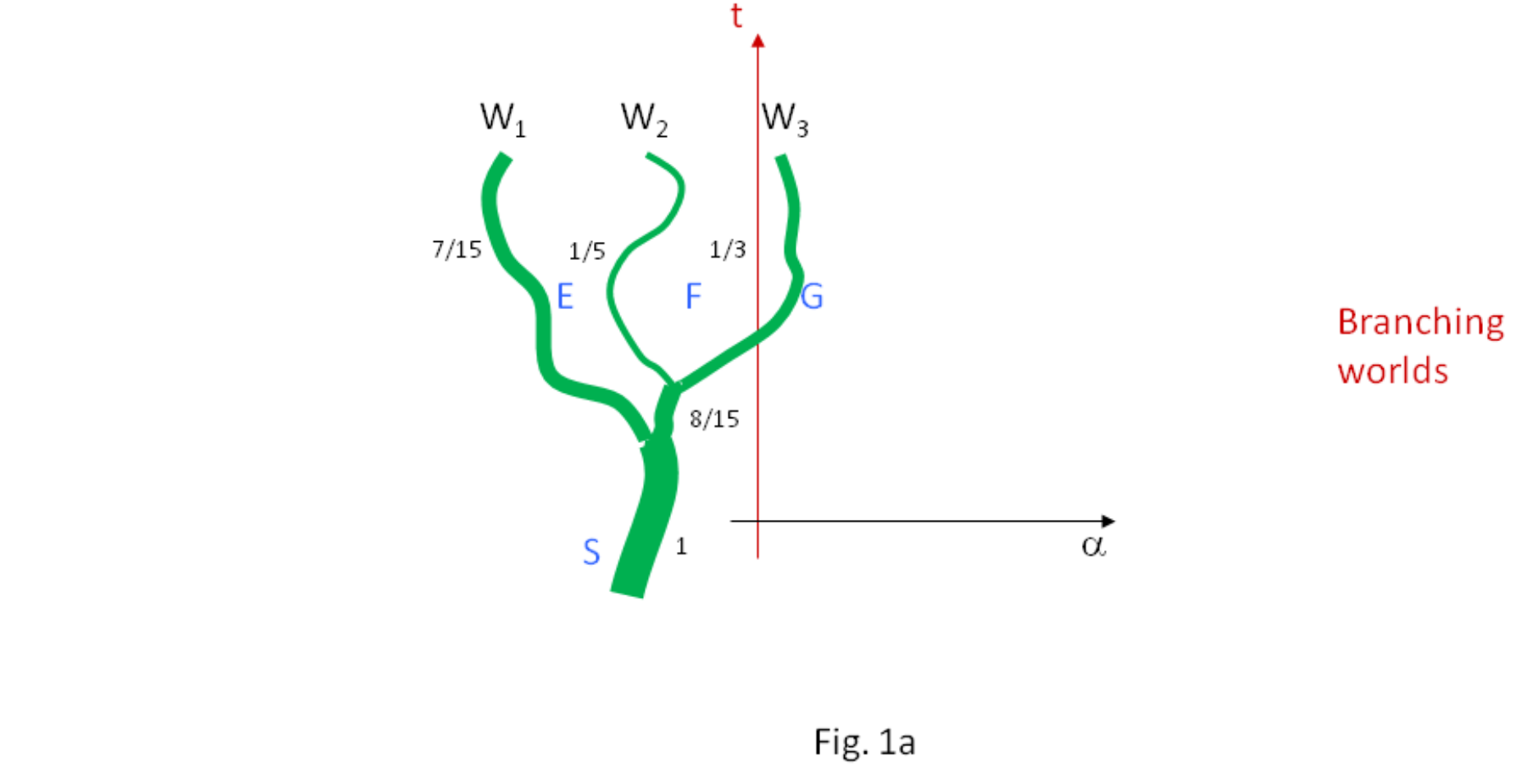}
\end{figure}
\begin{figure}
    \includegraphics[width=12cm, trim=0cm 10cm 2cm 10cm, clip=true ]{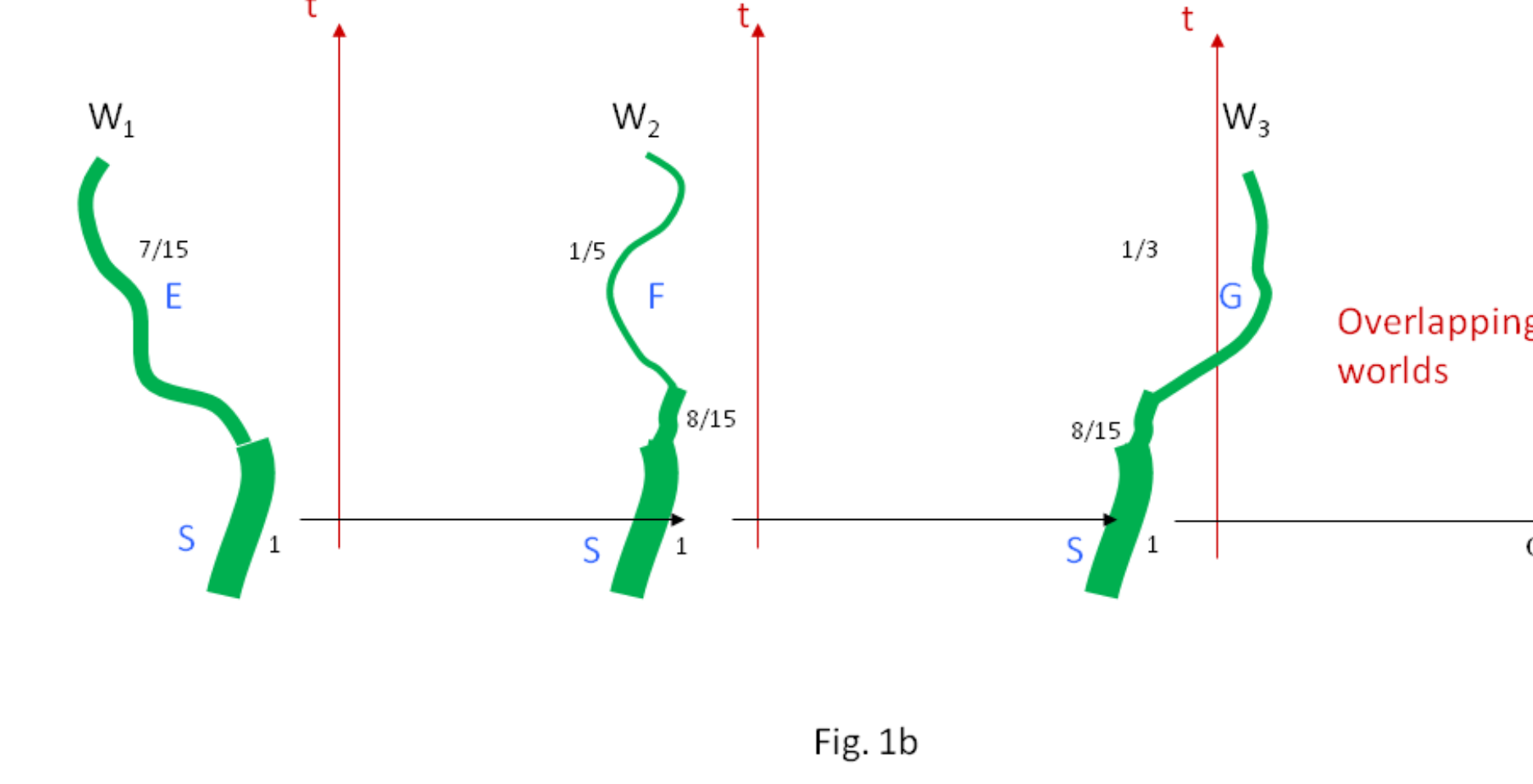}
\end{figure}
\begin{figure}
    \includegraphics[width=12cm, trim=0cm 10cm 2cm 10cm, clip=true ]{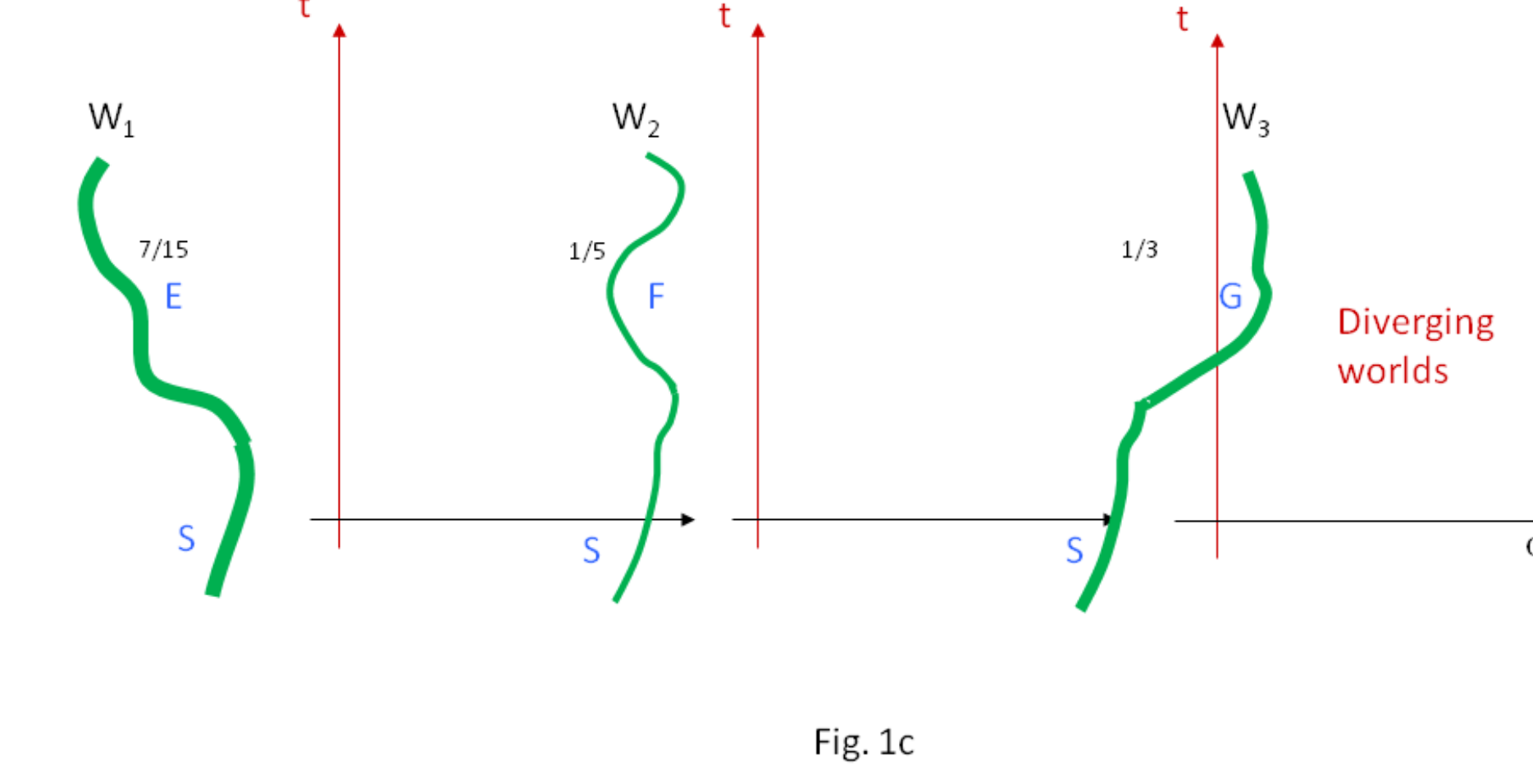}
\end{figure}

Here is the point made graphically. Fig 1a depicts the orbit of a Schr\"{o}dinger picture wave-function that develops into a superposition of components $E$, $F$ and $G$, with squared norms as indicated; this explains how branches came by their name. But the branches themselves can be depicted in
two ways, by either Fig 1b or Fig 1c. In Fig 1c it seems that branches do
not overlap, whereas in Fig 1b it seems that they do. Which of them is
correct?  Evidently the difference concerns only their amplitudes; of the
two, Fig 1c is the obvious representation for Heisenberg picture vectors,
which each have a unique amplitude. Fig 1b depicts rather graphs of
Schr\"{o}dinger picture states subject to state collapse, without renormalizing. Which of them is the `correct' picture?\ But once stated in this
way, the suspicion is that whether worlds in EQM diverge or overlap is
\textit{underdetermined} by the mathematics.\ One can use either picture; they are better
or worse adapted to different purposes.

If so, it is pretty clear which is the right one for making sense of
uncertainty. But of course Fig 1c is only a visual aid: can we be sure that
the Heisenberg-picture branch vectors do not overlap? This, perhaps, is a 
\textit{technical} philosophical question, to be settled by a
theory of parts and wholes for vectors -- by a vector mereology.

Alas, there is no such theory, or none on which there is any agreement.
Mereology, if it is close to any branch of mathematics, is close to set
theory (Lewis [1991)], but even there its links are controversial. We are
going to have to make this up as we go along. And it is a good question, at
this point, as to where the burden of proof really lies. Mereology has paid
little or no dividends in pure mathematics, let alone in physics. It is at
bottom an a priori metaphysical theory. A very good desideratum on any
reasonable metaphysics is that it makes sense of our best physical theory,
rather than nonsense.

Metaphysicians can, however, reasonably insist on an existence proof -- a demonstration
that at least one vector space mereology can be defined in terms of which
maximal branches in EQM do not overlap. Here is a simple construction that
depends essentially on the branching structure of a consistent history
space. Let maximal branch vectors $|\alpha \rangle $ be defined as before.
Any non-maximal branch vector $C_{\beta }|\Omega \rangle =|\beta \rangle $
can then be written (appendix 1):
\begin{equation*}
|\beta \rangle =\sum_{\alpha \in \beta }|\alpha \rangle .
\end{equation*}
By the consistency condition, $\langle \alpha |\alpha ^{\prime }\rangle =0$
for $\alpha \neq \alpha ^{\prime }$; therefore for any two branch vectors $%
|\beta \rangle ,$ $|\gamma \rangle $ if $\langle \beta |\gamma \rangle \neq
0 $ there exists a unique branch vector $|\delta \rangle $ and orthogonal
branch vectors $|\beta ^{\prime }\rangle ,|\gamma ^{\prime }\rangle $ such
that
\begin{equation*}
|\beta \rangle =|\beta ^{\prime }\rangle +|\delta \rangle ,\text{ }|\gamma
\rangle =|\gamma ^{\prime }\rangle +|\delta \rangle
\end{equation*}%
namely the vector:%
\begin{equation*}
|\delta \rangle =\sum_{\alpha \in \beta \cap \gamma }|\alpha \rangle .
\end{equation*}%
Our candidate mereology is then:\ a branch vector $|\epsilon \rangle $ is
part of $|\epsilon ^{\prime }\rangle )$ (denote $P(|\epsilon \rangle
,|\epsilon ^{\prime }\rangle $) if and only if either (a) $\epsilon
=\epsilon ^{\prime }$ or (b) there exists a non-zero branch vector $|\delta
\rangle $ such that $\langle \epsilon |\delta \rangle =0$, $|\epsilon
\rangle +|\delta \rangle =$ $|\epsilon ^{\prime }\rangle .$

Standard axioms of mereology are listed in appendix 2; it is easy to check
that they are satisfied by the parthood relation just defined. From the
overlap relation derived from $P$, we see that orthogonal vectors do not
overlap.

It follows that every branch vector is part of the universal state $|\Omega
\rangle $, as seems appropriate, but that orthogonal branch vectors have no
parts in common. Since, by consistency, branching always produces orthogonal
branch vectors, the branch vector representing Alice seeing spin down has no
part in common with that representing Alice seeing spin up. On this basis
branches of this kind diverge, in the philosophical sense, as do maximal
branches.

But in what sense, it might be asked, is \textit{Alice} represented as a part of a maximal branch
vector $|\alpha \rangle $?-- for the latter (on our candidate mereology) has
no vectors as proper parts. The answer is that Alice is represented not by a
vector at all, but by a sub-sequence $\beta $ of configurations of $\alpha
=\langle \alpha _{+},..,\alpha _{-}\rangle$ (see again $S1$, and our use of
ordered pairs $(\beta ,|\alpha \rangle )$ -- given which, questions about Alice and the part-whole relation in Alice's world are treated quasiclassically. It can hardly be objected that there must exist a
uniform theory of mereology that applies across the board to emergent
ontology. Far from it: that is a piece of metaphysics that has had little or
no success in the special sciences. We concede only that one might be
constructible in each domain.

With that it should be clear that our vector mereology on its own does not
settle the question of uncertainty. Alice's successors downstream of
branching diverge from each other, but they may still overlap with Alice
prior to branching, even given our vector mereology. A branch can be
coarse-grained from the time $t$ of branching in such a way that it is
degenerate with respect to Alice seeing spin-up and Alice seeing spin-down.
Thus a branch vector that represents only Alice's past $\alpha ^{-}(t)$ at $t
$ (of the form $|\alpha ^{-}(t)\rangle =P_{\alpha _{k}}(t_{k})...P_{\alpha
_{-}}(t_{-})|\Omega \rangle ,$ $t_{k}\leq t$, the Heisenberg-picture
analogue of Eq.(5)) represents Alice as containing all her possible future
selves, that is, as overlapping completely, according to our mereology, with
all her possible future selves, even if they do not overlap with each
other.

And there, in a nutshell, is the contrary view:\ by all means take the
parthood relation as specified, but let `Alice' be indexed to time, and be
represented by a Schr\"{o}dinger picture vector accordingly (as given by
Eq.(5)), or by the Heisenberg-picture branch vector $|\alpha ^{-}(t)\rangle $%
, rather than by relativization to a maximal branch. Alice in this sense
overlaps with all of her successors, and the grounds for self-locating
uncertainty evaporate. \ 

We always knew we had this option, however. This is to reject $S1$ not on
the grounds that worlds in EQM really overlap (on our proposed mereology
they do not), but because \textit{superpositions of worlds} overlap, and
because (so goes the objection) persons and things \textit{should} be
represented by superpositions of worlds, superpositions of vectors.

But why should they? We can do the same in the case of diverging worlds of
the sort usually considered by metaphysicians (having nothing to do with
quantum mechanics), or as they arise in cosmic inflation (see e.g. Tegmark [2010]. In either case it is uncontroversial that worlds do not
overlap, but still, \textit{sets} of worlds do. Let a person or a thing up
to time $t$ be a \textit{set} of worlds, namely of those that contain that
person or thing or one qualitatively identical up to time $t$ (which could
be far in the future): then there is no uncertainty of where a person or
thing is located among worlds that diverge after that time, either. But
metaphysicians are unlikely to take this proposal very seriously, not least because it makes nonsense of much of what we say. Nor should we in the
analogous case in EQM.\footnote{%
Note added Sep 2016. Paul Tappenden has argued for the merits of this picture on other grounds (Tappenden [2011a]. He also offers arguments why, even if there is no \textit{self-locating} uncertainty prior to branching in EQM, uncertainty can nevertheless be grounded on post-branching (Vaidman) uncertainty (Tappenden [2011b]). See also his criticisms of Saunders and Wallace [2008a,b] in Tappenden [2008]).} \bigskip

\noindent To conclude: there is no good reason to think EQM is really a
theory of overlapping worlds. If questions of overlap of branches are to be
settled by appeal to the underlying mathematics, in terms of vector space
structure, then there is at least one natural mereology in terms of which
worlds that differ in some feature, since orthogonal, are non-overlapping.
The semantics S1-S4 does \textit{not} misrepresent the underlying
mathematical structure of the theory; in terms of this mereology, it
correctly describes worlds as non-overlapping.

But it does follow that the word `branching' is something of a misnomer, in
the context of EQM, in the philosophers' sense of the word. But then the
word `diverging' is also infelicitous. `Branching' (inappropriately in EQM)
suggests overlap, `divergence' (inappropriately in EQM) suggests the absence
of physical relations. I see nothing wrong with continuing to call EQM a
theory of branching worlds, but only because the expression is
well-established among physicists\footnote{%
Although some talk of `parallel' worlds instead (see e.g. Tegmark [2010]). There is clearly a case for this terminology.} and because it is
fundamental to EQM that worlds, by superposition, do make up a dynamical
unity -- that they are all parts of the universal sate. At any rate, it will
mark a new phase in the status of the Everett interpretation if the debate
is over what the theory should be called.

\medskip

\noindent \textbf{Acknowledgements}: My debt to David Wallace is obvious,
but additional thanks are due to Harvey Brown, Hilary Greaves, and,
especially, Alastair Wilson, for helpful comments and suggestions.

\bigskip

\bigskip
\bigskip
\bigskip
\bigskip
\bigskip
\bigskip
\bigskip
\bigskip

\noindent {\Large Appendix 1: Decoherent quantum histories}

\bigskip

\noindent For each $k$, let $P_{\alpha_k}$, $\alpha_k =1,2,...$ be an
exhaustive, commuting set of projection operators on a Hilbert space $%
\mathcal{H}$, (a partition of unity):
\begin{equation*}
\sum_{\alpha_k}P_{\alpha_k }=I,\text{ }P_{\alpha_k }P_{\alpha_k^{\prime} }=\delta
_{\alpha_k \alpha_k^{\prime}}P_{\alpha_k }.
\end{equation*}%
The $\alpha_{k}$'s are to be thought of as coarse-grained cells of some parameter space (for the same of simplicity, say configuration space).\footnote{%
Coarse-grainings of phase space are also possible, even yielding projective-valued measures; for example, by using  von Neumann's construction (his `building bricks of reality', cited by Everett [1973]; von Neumann[1932 p.409]. Jumping foreward, see Halliwell [2010] for the definition of decoherent histories in terms of coarse grainings (integrals) of local densities.}
Let $H$ be the Hamiltonian, assumeed time-independent; the associated Heisenberg picture operators for time $t_k$ are then:
\begin{equation*}
P_{\alpha_k }(t_k)=e^{iHt_k/\hbar }P_{\alpha }e^{-iHt_k/\hbar}.
\end{equation*}%
The simplest history spaces are those for which the partition of unity is the same at each time. Let  $t_{1}<t_{2}<...<t_{n}$, and consider a history $\alpha=\langle\alpha_n,...,\alpha_1\rangle$; the associated chain (sometimes also called `class') operator is: 
\begin{equation*}
C_{\alpha }=P_{\alpha _{n}}(t_{n})P_{\alpha _{n-1}}(t_{n-1})...P_{\alpha
_{1}}(t_{1}).
\end{equation*}%
The $C_{\alpha }$'s are self-adjoint and positive (but not idempotent); they define a positive-operator-valued measure (POV measure). Acting on the state $|\Omega(t_0)\rangle $ at $t=0$ \ (abbreviate as $|\Omega\rangle$) we obtain branch state vectors $C_{\alpha }|\Omega\rangle=|\alpha\rangle .$ These are in 1:1 correspondence with Schr\"{o}dinger picture states $|\alpha(t_n)\rangle$, as would be obtained by measuring in sequence each $P_{\alpha_k}$ at time $t_k$, applying the projection postulate in each case. The correpondence (c.f. Eq.(5) section 4) is: 
\begin{equation*}
|\alpha(t_n)\rangle=e^{-iHt_n/\hbar }C_{\alpha}|\Omega\rangle.
\end{equation*}%
The probability $p(\alpha$ of history $\alpha$ is the product of the probabilities for each individual step (where the probability of each step is as given by the Born rule).  Formally:
\begin{equation}
p(\alpha)=\left\Vert |C_{\alpha }|\Omega \rangle \right\Vert
^{2}=Tr(C_{\alpha }\rho C_{\alpha }^{\dag })
\end{equation}%
where $\rho =|\Omega \rangle \langle \Omega |$ is the density matrix for the
state $|\Omega\rangle $ and `Tr' is the trace ($Tr(O)=\sum_{k}\langle
\phi _{k}|O\phi _{k}\rangle $, for any operator $O$ and orthonormal basis $%
\{|\phi _{k}\rangle\}$ over $\mathcal{H}$). Likewise, the conditional
probability of $\alpha $ (for $t_{n}<...<t_{k+1}$) given $\beta $ (for $%
t_{k}<...<t_{1}$) is%
\begin{equation*}
p_{\rho }(\alpha /\beta )=\frac{Tr(C_{\alpha \ast \beta }\rho C_{\alpha \ast
\beta }^{\dag })}{Tr(C_{\beta }\rho C_{\beta }^{\dag })}.
\end{equation*}%
where $\alpha \ast \beta $ is the history comprising $\beta $ (up to time $%
t_{k}$) and $\alpha $ (from $t_{k+1}$ to $t_{n}$). But whether or not these quantities can really be interpreted as probabilities -- granted that repeatable measurements of the projectors $P_{\alpha_k}$ are \textit{not} in fact performed -- is another matter. 

Consider again the concept of coarse-graining of a parameter space (like configuration
space). It extends naturally to chain operators: for each $k$, let $\{\overline{\alpha_k}%
\}$ be a coarse graining of $\{\alpha_k\}$, so that each finer-grained cell $%
\alpha_k$ is contained in some coarser grained cell $\overline{\alpha_k}$ in
the parameter space. We can then speak of coarsening and fine-grainings of
histories too. Now consider a set of histories with chain operators $%
\{C_{\alpha }\}$, and a coarse-graining with chain operators $\{C_{\overline{%
\alpha }}\}.$ The two are simply related:%
\begin{equation*}
C_{\overline{\alpha }}=\sum_{\alpha \in \overline{\alpha }}C_{\alpha }
\end{equation*}%
where the sum is over all finer-grained histories $\alpha $ contained within 
$\overline{\alpha }$. However, it by no means follows that the i probabilities of coarse-grainings of fine-grained histories are the sum of the probabilities of the fine-grained histories. The \textit{sum rule}
\begin{equation}
p(\overline{\alpha })=\sum_{\alpha \in \overline{\alpha }}p(\alpha ).
\end{equation}%
is a substantive constraint.\footnote{%
For example, in the 2-slit experiment, it is not satisfied by histories so fine-grained as to show through which slit each particle goes.}
 A sufficient condition for the sum rule is that vectors representing the fine-grained histories are approximately orthogonal:%
\begin{equation}
\langle C_{\alpha }\Omega|C_{\alpha ^{\prime }}\Omega\rangle=\langle\alpha|\alpha^{\prime}\rangle\approx0\text{,  }\alpha \neq\alpha ^{\prime }.
\end{equation}%
 Histories (for fixed Hamiltonian $H$ and state $|Omega\rangle$ that satisfy (8) are called \textit{consistent} (by Griffiths and Omn\`{e}s); 
\textit{(medium) decoherent} (by Gell-Mann and Hartle)\footnote{%
The necessary and sufficient condition for the sum rule (7) is that only the real part of (9) vanish (or approximately vanish). But the stronger condition is the more robust, and is automatically satisfied in a quasiclassical domain.} Given consistency, Everett's relativization is a transitive relation even in time-like
directions; it is automatically transitive in space-like
directions by virtue of microcausality).

The additivity condition for a 1-place history, assuming (6),  is automatically satisfied (Everett in his [1957] turned this reasoning around:\ assuming additivity, he derived (6)). It is
satisfied by two-time histories as well, but in the general case it fails. A maximal set of consistent histories defines a \textit{consistent history space}. If in addition the histories of such a space obey quasi-classical equations, the history space is called (following Gell-Man and Hartle [1993]) a \textit{quasiclassical domain}. 

It follows too that for any consistent history space there exists a
fine-graining $\{P_{\alpha }\}$ which is consistent and for which, for any $%
t_{n}>t_{m}$ and for any $\alpha _{n}$ with $P_{\alpha _{n}}(t_{n})|\Psi
\rangle \neq 0$, there exists exactly one $\alpha _{m}$ such that 
\begin{equation*}
P_{\alpha _{n}}(t_{n})P_{\alpha _{m}}(t_{m})|\Psi \rangle \neq 0
\end{equation*}%
(Griffiths [1993], Wallace [2012 p.92-94.]). That is, for each $\alpha _{n}$ at time $%
t_{n}$, there is a \textit{unique} history preceding it -- the set of
histories can be fine-grained so as to have a purely branching structure
(with no recombination of branches). 

The consistency condition and the quantum histories formalism is widely
considered a generalization of quantum theory as,
fundamentally, a theory of probability. As such there is a continuum
infinity of consistent history spaces available -- new resources for the
exploration of quantum systems, indeed. But from the point of view of
EQM, consistency is far too weak a condition to give
substance to the notion of histories as autonomous and robust 
structures in the universal sate, our abiding criteria for the existence of worlds. 
\bigskip
\bigskip

\noindent {\Large Appendix 2: Axioms of mereology}

\bigskip

\noindent We write `$Pxy$' for `$x$ is part of $y$'; this relation is
reflexive, transitive, and antisymmetric. Defined notions are:

\begin{description}
\item[Overlap] $Oxy\underset{def}{=}\exists z(Pzx\&Pzy).$

\item[Underlap] $Uxy\underset{def}{=}\exists z(Pxz\&Pyz).$

\item[Fusion] $x\sqcup y\underset{def}{=}\iota z\forall
w(Pxw\&Pyw\rightarrow Pzw).$
\end{description}

\noindent (In words, $x$ and $y$ overlap if a part of $x$ is a part of $y$;
they underlap if there is something of which they are both parts; and their
fusion is the unique thing that is a part of anything of which $x$ and $y$
are both parts.) The elementary axioms of mereology are:

\begin{description}
\item[M1] $\forall x(x\sqcup x)=x$

\item[M2] $\forall x\forall y(Uxy\rightarrow \exists z(z=x\sqcup y))$

\item[M3] $\forall z(Pxz\&Pyz\rightarrow Px\sqcup yz)$
\end{description}

\noindent (the fusion of anything with itself is itself; if two things
underlap then their fusion exists; anything which underlaps two things is
part of the fusion of those things). Define the parthood relation as above:

\begin{description}
\item[Vector Part] $P(|\beta \rangle ,|\gamma \rangle )$ if and only if
either (a) $\beta =\gamma ,$ or (b) there exists a non-zero branch $|\delta
\rangle $ such that $\langle \delta |\beta \rangle =0$, $|\beta \rangle
+|\delta \rangle =$ $|\gamma \rangle .$
\end{description}

It is simple to check that $P$ is reflexive, transitive, and antisymmetric.
The fusion operation as defined by $P$ is:

\begin{itemize}
\item If $\langle \beta |\gamma \rangle =0,|$ $\beta \rangle \sqcup |\gamma
\rangle =|\beta \rangle +|\gamma \rangle $

\item If $\langle \beta |\gamma \rangle \neq 0,$ $\beta \in \gamma $, $|$ $%
\beta \rangle \sqcup |\gamma \rangle =|\gamma \rangle $; $\gamma \in \beta
,| $ $\beta \rangle \sqcup |\gamma \rangle =|\beta \rangle .$
\end{itemize}

\noindent It clearly satisfies $M1$-$M3.$

\bigskip \
\begin{center}
{\LARGE References\bigskip }
\end{center}

\setlength{\parindent}{-0.7cm} Aharonov, Y., D. Albert, and L. Vaidman [1988], `How the result of
a measurement of a component of the spin of a spin-$\frac{1}{2}$ particle can turn out to be 100', \textit{Physical Review Letters }\textbf{60%
}, 1351-54.

\setlength{\parindent}{-0.7cm} Aharanov, Y., and L. Vaidman [1993], `Protective measurements', 
\textit{Physics Letters} \textbf{A 178} p. 38.

\setlength{\parindent}{-0.7cm} Davidson, D. [2001], \textit{Inquiries into Truth and
Interpretation} (2nd ed.). Oxford University Press.

\setlength{\parindent}{-0.7cm}Deutsch, D. [1999], `Quantum theory of probability and decisions', 
\textit{Proceedings of the Royal Society of London} \textbf{A455},
3129--3137. Available online at http://arxiv.org/abs/quant-ph/9906015.

\setlength{\parindent}{-0.7cm}--- [2010], 'Apart from universes', in Saunders et al [2010], pp.542-52.

\setlength{\parindent}{-0.7cm}Everett III, H. [1957], `Relative state formulation of quantum
mechanics', \textit{Reviews of Modern Physics} \textbf{29}, 454-62.

--- [1973], `Theory of the universal wave-function',
in \textit{The Many-Worlds Interpretation of Quantum Mechanics}, B. De Witt and N. Graham (eds.), pp.3-140, Princeton University Press. 

\setlength{\parindent}{-0.7cm}Gell-Mann, M. and J.B. Hartle [1990], `Quantum Mechanics in the
light of quantum cosmology', in\textit{\ Complexity, Entropy, and the
Physics of Information}, W.H. Zurek, ed., Reading, Addison-Wesley.

---  [1993], `Classical equations for
quantum systems', \textit{Physical Review D} \textbf{47}, 3345-382.
Available online at http://arxiv.org/abs/gr-qc/9210010.

\setlength{\parindent}{-0.7cm} Greaves, H. [2004], `Understanding Deutsch's probability in a
deterministic multiverse', \textit{Studies in History and Philosophy of
Modern Physics} \textbf{35}, 423-56. Available online at
http://philsci-archive.pitt.edu/archive/00001742/.

---  [2007], `On the Everettian epistemic problem', \textit{%
Studies in History and Philosophy of Modern Physics \textbf{38}, }120-152.
Available online at http://philsci-archive.pitt.edu/archive/00002953.

\setlength{\parindent}{-0.7cm} Greaves, H. and W. Myrvold [2010], `Everett and evidence', in Saunders et al [2010], pp.264-304.

\setlength{\parindent}{-0.7cm} Griffiths, R. [1993], `Consistent interpretation of quantum
mechanics using quantum trajectories', \textit{Physical Review Letters} 
\textbf{70}, 2201-2204.

\setlength{\parindent}{-0.7cm} Halliwell, J. [2010], `Macroscopic superposition, decoherent histories, and the emergence of hydrodynamic behaviour', in Saunders et al [2010], pp.99-117.

\setlength{\parindent}{-0.7cm} Ismael, J. [2003], `How to combine chance and determinism:
Thinking about the future in an Everett universe', \textit{Philosophy of
Science} \textbf{70}, 776--790.

\setlength{\parindent}{-0.7cm} Johnston, M. [1989], `Fission and the facts', \textit{%
Philosophical Perspectives}, \textbf{3}, 369-397.

\setlength{\parindent}{-0.7cm} Kent, A. [2010], `One world versus many: the inadequacy of
Everettian accounts of evolution, probability, and scientific confirmation',
in Saunders et al [2010].

\setlength{\parindent}{-0.7cm} Ladyman, J. and D. Ross [2007], \textit{Every Thing Must Go:
Metaphysics Naturalized, }Oxford.

\setlength{\parindent}{-0.7cm} Lewis, D. [1973], \textit{Counterfactuals}. Harvard.

--- [1986a], \textit{Philosophical Papers, Vol. 2}, Oxford:
Oxford.

--- [1986b], \textit{On the Plurality of Worlds}, Blackwell.

--- [1991], \textit{Parts of Classes}, Blackwell.

\setlength{\parindent}{-0.7cm} Lubkin, E. [1979], `An application of ideal experiments to quantum mechanical measurement theory', \textit{International Journal of Theoretical Physics} \textbf{18}, 165-77.

\setlength{\parindent}{-0.7cm} Papineau, D. [1996], `Comment on Lockwood', \textit{British
Journal for the Philosophy of Science} \textbf{47}, 233-41.

\setlength{\parindent}{-0.7cm} Saunders, S. [1994], `Decoherence and evolutionary adaptation', 
\textit{Physics Letters} \textit{A}\textbf{\ 184}, p.1-5.

--- [1997], `Naturalizing metaphysics', \textit{The Monist} \textbf{80}, 44-69.

--- [1998], `Time, quantum mechanics, and probability', 
\textit{Synthese}, \textbf{114}, pp.405-44. Available online at
http://arxiv.org/abs/quant-ph/0112081.

--- [2004], `Derivation of the Born Rule from Operational
Assumptions', \textit{Proceedings of the Royal Society} \textbf{A 460}, 1-18. Available online at https://arxiv.org/abs/quant-ph/0211138

--- [2005], `What is probability?', in \textit{Quo Vadis
Quantum Mechanics}, A. Elitzur, S. Dolev, and N. Kolenda, eds., Springer.
Available online at http://arxiv.org/abs/quant-ph/0412194.

\setlength{\parindent}{-0.7cm} Saunders, S., and D. Wallace [2008a] `Branching and uncertainty', 
\textit{British Journal for the Philosophy of Science} \textbf{59}, 293 -
305. Available online at http://philsci-archive.pitt.edu/archive/00003811/.

--- [2008b], `Saunders and Wallace
reply', \textit{British Journal for the Philosophy of Science} \textbf{59},
315-17.

\setlength{\parindent}{-0.7cm} Saunders, S., J. Barrett, A. Kent, and D. Wallace (eds.)[2010], \textit{%
Many Worlds? Everett, Quantum Theory, and Reality}, Oxford.

\setlength{\parindent}{-0.7cm} Savage, L. [1954], \textit{The Foundations of Statistics}, Wiley.

\setlength{\parindent}{-0.7cm} Tappenden, P. [2008], `Comment on Saunders and Wallace', \textit{
British Journal for the Philosophy of Science} \textbf{59}, 306-314.

--- [2011a], `A metaphysics for semantic internalism', \textit{Metaphysica}\textbf{12}, 125-36. Available online at http://philsci-archive.pitt.edu/8721/.

-- [2011b], ‘Evidence and uncertainty in Everett’s multiverse’, \textit{British
Journal for the Philosophy of Science} \textbf{62}, pp. 99–123. Available online at http://philsci-archive.pitt.edu/5046/.

\setlength{\parindent}{-0.7cm} Tegmark, M. [2010], 'Many worlds in context', in Saunders et al [2010] pp.553-81.

\setlength{\parindent}{-0.7cm} Uffink, J. [2000] `How to protect the interpretation of the wave
function against protective measurements'. Available online at
http://arxiv.org/abs/quant-ph/9903007v1.

\setlength{\parindent}{-0.7cm} Vaidman, L. [1998], `On schizophrenic experiences of the neutron
or why we should believe in the many-worlds interpretation of quantum
theory', \textit{International Studies in the Philosophy of Science} \textbf{%
12}, 245--261.

\setlength{\parindent}{-0.7cm} Van Fraassen, B. [1980], \textit{The Scientific Image}, Oxford:
Clarendon Press.

\setlength{\parindent}{-0.7cm} Von Neumann, J. [1932], \textit{Mathematische Grundlagen Der
Quantenmechanik}, translated by R.T. Beyer as \textit{Mathematical
Foundations of Quantum Mechanics}, Princeton University Press, 1955.

\setlength{\parindent}{-0.7cm} Wallace, D. [2005], `Language use in a branching universe',
available online at http://philsci-archive.pitt.edu/archive/00002554.

---  [2006], `Epistemology quantized: circumstances in
which we should come to believe in the Everett interpretation', \textit{%
British Journal for the Philosophy of Science }\textbf{57}, 655-689.
Available online at http://philsci-archive.pitt.edu/archive/00002839.

--- [2010], `How to prove the Born rule', in Saunders et
al [2010] pp.227-63. Available online at http://philsci-archive.pitt.edu/4709/

--- [2012], \textit{The Emergent Multiverse: Quantum Theory according to the Everett Interpretation}, Oxford.

\setlength{\parindent}{-0.7cm} Wilson, A. [2011], `Macroscopic ontology in Everettian quantum mechanics’,
\textit{Philosophical Quarterly} \textbf{ 61}, pp. 363–82.

--- [2013], `Objective probability in Everettian quantum mechanics', \textit{Britisch Journal for the Philosophy of Science} \textbf{64}, 709-37.

\setlength{\parindent}{-0.7cm} Zurek, W. [2005], `Probabilties from entanglement, Born's rule $p_k=|\psi_k|^2$ from envariance', \textit{Physical Review} \textbf{A71}, 052105. Available online at https://arxiv.org/abs/quant-ph/0405161

\bigskip

\end{document}